\documentclass[onecollarge]{svjour2}       
\usepackage{graphicx}
\usepackage{amsmath}
\usepackage{cite}
\journalname{}
\begin{document}

\title{The stationary Dirac equation as a generalized Pauli equation for two quasiparticles}


\author{Nikolay L. Chuprikov 
}


\institute{N. L. Chuprikov \at
              Tomsk State Pedagogical University, 634041, Tomsk, Russia \\
              \email{chnl@tspu.edu.ru}           
}

\date{Received: date / Accepted: date}

\maketitle

\begin{abstract}
By analyzing the Dirac equation with static electric and magnetic fields it is shown that Dirac's theory is nothing but a generalized one-particle
quantum theory compatible with the special theory of relativity. This equation describes a {\it quantum} dynamics of a {\it single} relativistic
fermion, and its solution is reduced to solution of the generalized Pauli equation for two quasiparticles which move in the Euclidean space with
their effective masses holding information about the Lorentzian symmetry of the four-dimensional space-time. We reveal the correspondence between
the Dirac bispinor and Pauli spinor (two-component wave function), and show that all four components of the Dirac bispinor correspond to a fermion
(or all of them correspond to its antiparticle). Mixing the particle and antiparticle states is prohibited. On this basis we discuss the
paradoxical phenomena of Zitterbewegung and the Klein tunneling.

\end{abstract}

\newcommand{\ppp}{\mbox{\hspace{5mm}}}
\newcommand{\ooo}{\mbox{\hspace{3mm}}}
\newcommand{\ooa}{\mbox{\hspace{1mm}}}
\newcommand{\ppd}{\mbox{\hspace{18mm}}}
\newcommand{\ppt}{\mbox{\hspace{34mm}}}
\newcommand{\ppo}{\mbox{\hspace{10mm}}}
\newcommand{\lcom}{\lambda\hspace{-1mm}\bar{}\hspace{1mm}}

\section{Introduction}

As known (see, e.g., \cite{Bar}), the elaboration of a fully consistent interpretation of Dirac's equation, as a relativistic analog of
Schr\"{o}dinger's one-particle equation, represents one of the most fundamental and difficult problems of theoretical physics. The relationship
between the (Lorentzian) Dirac formalism and the (Galilean) Schr\"{o}dinger dynamics is nontrivial even in the non-relativistic limit \cite{Hol}.
In the general case, because of paradoxical physical implications of the Dirac equation, its interpretation as a quantum-mechanical equation for a
one-particle wave function faces serious problems. As a consequence, so far there is a widespread opinion that this equation cannot be considered
as the generalization of the Schr\"{o}dinger-Pauli equation for single relativistic fermions; Dirac theory is rather considered (see, e.g.,
\cite{Mess}) as a field theory which is in need of quantization.

Perhaps, the most paradoxical implications of the Dirac equation are the Klein tunneling \cite{Mess,Hans,Hols,Domb,Bos,Kon,Alha,Iran} and the
so-called 'Zitterbewegung' phenomenon \cite{Mess,Ger,Con}. Both are often mentioned in the current literature on this equation and both give rise
to controversy among researchers. Even the very nature of both these phenomena is differently understood in the current literature. For example,
the Klein paradox for an electron scattering on a strong electric scalar step potential (when its energy lies, on the energy scale, below the step
height in the so-called Klein zone) is understood by some authors as the appearance of a classically accessible region behind the step; at the
same time others talk about the Klein paradox when, in this scattering problem, the probability flow associated with reflected particles exceeds
the incident flow.

Regarding Zitterbewegung, there are at least two main, principally different versions of this phenomenon, Schr\"{o}dinger's \cite{Schr} and
Hestenes' \cite{Hest} ones. The latter treats an electron as "a rapidly rotating electric dipole"\/, i.e., as a particle having an internal
structure. This model of Zitterbewegung needs a reformulation of the standard Dirac equation, and what is more important is that it, in fact, lies
beyond the scope of quantum mechanics itself. It can be considered as a {\it pre-quantum} Zitterbewegung model. And, since our final task is to
study this phenomenon from the quantum-mechanical point of view, we shall consider only the Schr\"{o}dinger version \cite{Schr} of Zitterbewegung,
whereby this phenomenon follows from the standard Dirac equation when one assumes that a particle might be in a quantum state representing the
superposition of the particle and antiparticle states (the electron itself is considered as a point object).

The idea that such states may coexist with each other underlies also most approaches of the Klein tunneling, whereby this phenomenon is treated as
a many-particle effect accompanied by creation of electron-positron pairs. This interpretation lies obviously beyond the scope of Dirac theory, as
quantum formalism that describes one particle. At the same time this equation remains valid at all energies of a relativistic particle (see, e.g.,
\cite{Grei,Bjor}) and hence there is no reason to discard the possibility to interpret the Klein tunneling as a single-particle phenomenon.

Instead of the pair-creation mechanism, some authors (see, e.g., \cite{Bos,Alha}) attempt to resolve the Klein paradox by making use of different
kinds of 'ghost' wave modes and virtual particles. In these approaches the probability flow associated with 'ghost' modes and virtual particles
balances the electron flow at the step, and the Klein tunneling disappears. But this result, too, cannot be considered as satisfactory, because it
contradicts the studies of the Klein tunneling in graphene where this effect, predicted on the basis of the Dirac equation, really exists (see,
e.g., \cite{Been}).

We consider that the source of all paradoxes that surround at present the Dirac equation is the existing practice to associate the 'small'
component of the Dirac bispinor with an antiparticle (see, e.g., \cite{Mess}). Our aim is to show that this practice, based on the assumption that
the particle and antiparticle quantum states belong to the same Hilbert space, is unfounded. We present Dirac theory as a generalized
Schr\"{o}dinger-Pauli formalism that describes the dynamics of single relativistic fermions. On this basis we discuss the above paradoxical
phenomena -- the Klein tunneling and Zitterbewegung.

\section{Dirac dynamics as a generalized Schr\"{o}dinger dynamics of two spin-1/2 quasiparticles with different effective masses}

Let us consider the (3+1)-Dirac equation with the static electric scalar potential $\phi(\vec{r})$ and vector potential $\vec{\cal{A}}(\vec{r})$:
\begin{eqnarray} \label{1}
\left[\left(
\begin{array}{cc}
\textbf{1} & 0 \\
0 & -\textbf{1}
\end{array} \right)\left(i\hbar\frac{\partial}{\partial t}-V(\vec{r})\right)+ \left(
\begin{array}{cc}
0 & \vec{\sigma} \\
-\vec{\sigma} & 0
\end{array} \right)\left(i\hbar c\vec{\nabla}+e\vec{\cal{A}}\right)-mc^2\right]\left(
\begin{array}{cc}
\Phi \\
\chi
\end{array}\right)=0
\end{eqnarray}
where $V=e\phi$; $e$ is the electric charge of a particle; $m$ is its (rest) mass; $\vec{r}$ is its radius-vector; $\sigma^1$, $\sigma^2$ and
$\sigma^3$ are the Pauli matrices. The corresponding continuity equation is
\begin{eqnarray} \label{2}
\frac{\partial W}{\partial t}+\vec{\nabla}\vec{J}=0;\ppp W=|\Phi|^2+|\chi|^2,\ppp \vec{J}=c(\Phi^*\vec{\sigma}\chi+\chi^*\vec{\sigma}\Phi).
\end{eqnarray}

According to the current vision of Eq. (\ref{1}) (see, e.g., \cite{Mess}), each component of the Dirac bispinor is associated with a given
orientation of the spin along the axis $OZ$ and with a given sign of the particle energy: if $\Phi=\left(
\begin{array}{cc}
\Phi_+ \\
\Phi_-
\end{array}\right)$ and $\chi=\left(
\begin{array}{cc}
\chi_+ \\
\chi_-
\end{array}\right)$, then the pairs $(\Phi_+,\chi_+)$ and $(\Phi_-,\chi_-)$ describe a particle whose
$z$-projection of spin is $+1/2$ and $-1/2$, respectively. As regards the spinors $\Phi$ and $\chi$ (referred to, in the non-relativistic limit,
as 'large' and 'small' components, respectively), they are assumed to correspond to the positive and negative values of the particle energy
\cite{Mess}.

However, there is every reason to believe that this is not the case. Indeed, since the scalar and vector potentials are static, one can search a
particular (stationary) solution of Eqs. (\ref{1}) in the form
\begin{eqnarray} \label{3}
\left(\begin{array}{cc}
\Phi(\vec{r},t) \\
\chi(\vec{r},t)
\end{array}\right)=\left(\begin{array}{cc}
\Phi(\vec{r};E) \\
\chi(\vec{r};E)
\end{array}\right) e^{-iEt/\hbar}
\end{eqnarray}
where $E$ is the particle energy. For the stationary state we have
\begin{eqnarray} \label{4}
(E-V-mc^2)\Phi+(i\hbar c\vec{\nabla}+e\vec{\cal{A}})\vec{\sigma}\chi=0,\ppp (E-V+mc^2)\chi+(i\hbar
c\vec{\nabla}+e\vec{\cal{A}})\vec{\sigma}\Phi=0;
\end{eqnarray}

Let $\epsilon=E-mc^2$, ${\cal{D}}=\vec{\sigma}\vec{\nabla}-\frac{ie}{\hbar c}\vec{\sigma}\vec{\cal{A}}$. Then Eqs. (\ref{4}) and the expressions
from Eqs. (\ref{2}) for $W$ and $\vec{J}$ can be rewritten as
\begin{subequations}
\begin{align}
-\frac{\hbar^2}{2}{\cal{D}}\frac{1}{M}{\cal{D}}\Phi+V\Phi=\epsilon\Phi;\ppp \chi=-\frac{i\hbar}{2Mc}{\cal{D}}\Phi;
\ppp M(\epsilon,V)=m\left(1+\frac{\epsilon-V}{2mc^2}\right) \label{5a}\\
W=|\Phi|^2+\frac{\hbar^2}{4M^2c^2}|{\cal{D}}\Phi|^2,\ppp
\vec{J}=\frac{i\hbar}{2M}\left[({\cal{D}}\Phi)^*\vec{\sigma}\Phi-\Phi^*\vec{\sigma}{\cal{D}}\Phi\right]. \label{5b}
\end{align}
\end{subequations}
Note that the set of Eqs. (\ref{5a}) is exactly equivalent to Eqs. (\ref{4}). Thus, solving the Dirac equation is reduced, in fact, to solving Eq.
(\ref{5a}) for the spinor $\Phi$, which represents the generalized Schr\"{o}dinger equation for a quasiparticle with the effective mass $M$.
Considering that $\vec{H}=[\vec{\nabla}\times\vec{\cal{A}}]$ is the magnetic field and
\begin{eqnarray} \label{6}
{\cal{D}}^2=\left(\vec{\nabla}-\frac{ie}{\hbar c}\vec{\cal{A}}\right)^2+\frac{e}{\hbar c}\vec{\sigma}\vec{H},
\end{eqnarray}
we rewrite this equation as a generalized Pauli equation for a quasiparticle with the effective mass $M$:
\begin{eqnarray} \label{7}
-\frac{\hbar^2}{2M}\left[\left(\vec{\nabla}-\frac{ie}{\hbar c}\vec{\cal{A}}\right)^2 -\frac{\vec{\sigma}\vec{\nabla}M}{M}\cdot{\cal{D}}\right]\Phi
+\left(V-\frac{e\hbar}{2Mc}\vec{\sigma}\vec{H}\right)\Phi=\epsilon\Phi.
\end{eqnarray}

As known, equations of such a kind (without the vector potential $\vec{\cal{A}}$) play an essential role in solid state physics (see, e.g.,
\cite{Burt,Kar}), where they describe the quantum dynamics of a Bloch electron in superlattices. Specifically, they arise within the
effective-mass approximation as equations for the envelope of the wave function of a Bloch electron. In this approximation, the effective mass of
this quasiparticle, in each layer of a superlattice, carries information about the periodic potential in the layer. And, according to this
approach, the envelop of the wave function must be everywhere continuous together with its first spatial derivative divided by the effective mass.

Essentially the same situation arises for a Dirac particle. Now, to ensure the continuity of the probability density $W$ and the probability
current density $\vec{J}$ (\ref{5b}) at the points where the scalar and vector potentials are discontinuous, the spinor $\Phi$ must be everywhere
continuous together with the spinor $M^{-1}{\cal{D}}\Phi$. Obviously that the last requirement is also referring to the continuity of the spinor
$\chi$. Like the effective mass of a Bloch electron, the effective mass $M$ associated with the spinor $\Phi$ keeps information about the
Lorentzian symmetry of the four dimensional space-time. However, unlike the equation for the envelope of the wave function of a Bloch electron,
Eq. (\ref{5a}) (or Eq. (\ref{7})) is exact.

Note that the probability density $W$ is determined not only by the term $|\Psi|^2$. The expression (\ref{5b}) for this quantity contains also the
term proportional to $|{\cal{D}}\Phi|^2$. Of course, being associated with the spinor $\chi$, it vanishes in the non-relativistic limit when
$\epsilon,\ooa V,\ooa e\vec{\cal{A}}\ll mc^2$. This fact is commonly taken (see, e.g., p.934 in \cite{Mess}) as a good cause for neglecting the
spinor $\chi$ in this limit. But this is mistaken in principle: the validity of the inequality $|\chi|^2\ll |\Phi|^2$ does not at all mean that
the 'small' component $\chi$ is inessential in this limit, in comparison with the 'large' component $\Phi$. Firstly, we have to recall that the
second-order differential equation (\ref{5a}) is equivalent to the system (\ref{4}) of coupled first-order differential equations for $\Phi$ and
$\chi$, where both these components are equally important. Secondly, both are also equally important in expression (\ref{2}) for the probability
current density (see also Section \ref{concl}).

To elucidate the role of the 'small' component $\chi$ it is useful to express Eqs. (\ref{5a}) and (\ref{5b}) in another equivalent form, where
$\Phi$ and $\chi$ change roles:
\begin{subequations}
\begin{align}
-\frac{\hbar^2}{2}{\cal{D}}\frac{1}{\mu}{\cal{D}}\chi+(V-2mc^2)\chi=\epsilon\chi;\ppp \Phi=-\frac{i\hbar}{2\mu c}{\cal{D}}\chi;
\ppp \mu(\epsilon,V)=\frac{\epsilon-V}{2c^2} \label{8a}\\
W=\frac{\hbar^2}{4\mu^2c^2}|{\cal{D}}\chi|^2+|\chi|^2,\ppp
\vec{J}=\frac{i\hbar}{2\mu}\left[({\cal{D}}\chi)^*\vec{\sigma}\chi-\chi^*\vec{\sigma}{\cal{D}}\chi\right]. \label{8b}
\end{align}
\end{subequations}
Thus, solving Eqs. (\ref{4}) reduces now to solving Eq. (\ref{8a}) for the spinor $\chi$, and the analog of Eq. (\ref{7}) is
\begin{eqnarray} \label{77}
-\frac{\hbar^2}{2\mu}\left[\left(\vec{\nabla}-\frac{ie}{\hbar c}\vec{\cal{A}}\right)^2
-\frac{\vec{\sigma}\vec{\nabla}\mu}{\mu}\cdot{\cal{D}}\right]\chi +\left(V-2mc^2-\frac{e\hbar}{2\mu c}\vec{\sigma}\vec{H}\right)\chi=\epsilon\chi.
\end{eqnarray}
This equation represents the generalized Pauli equation for the quasiparticle that has the effective mass $\mu$ and moves, as a heavy
quasiparticle, in the same vector potential but in a scalar potential reduced now by $2mc^2$. In this case the spinor $\chi$ and also
$\mu^{-1}{\cal{D}}\chi$ must be continuous. The last condition guarantees the continuity of $\Phi$; hence $W$ and $\vec{J}$ will be continuous
too.

So, the quantum ensemble of a Dirac particle with the energy $E$ consists of two subensembles: the subensemble of 'heavy' spin-1/2 quasiparticles
with effective mass $M$ and the subensemble of 'light' spin-1/2 quasiparticles with the effective mass $\mu$; in this case $M-\mu=m$ and
$M+\mu=(E-V)/c^2\equiv \cal{M}$. Such partitioning of the original ensemble of a Dirac particle is unique, because no effective mass can be
assigned to any superposition $c_1\Phi+c_2\chi$ with $c_1\neq 0$ and $c_2\neq 0$. Only the (stationary) spinors $\Phi$ and $\chi$ by themselves,
arising within the standard representation of the Dirac equation, can be associated with quasiparticles possessing definite effective masses.

Thus, according to our approach, in the static external potentials $V$ and $\vec{\cal{A}}$, a Dirac particle moves, with the probability
$|\Phi|^2$, just as a Schr\"{o}dinger spin-1/2 quasiparticle moving in these fields with the effective mass $M$. And, with the probability
$|\chi|^2$, it moves just as a Schr\"{o}dinger spin-1/2 quasiparticle with the effective mass $\mu$ moves in the same vector potential but in the
reduced scalar potential $V-2mc^2$. In fact, the effective masses $M$ and $\mu$ must be considered, together with the spin projections $+\hbar/2$
and $-\hbar/2$, as quantum numbers characterizing the components of the (stationary) Dirac bispinor.

It is of importance is to stress once again that the generalized Pauli equations (\ref{7}) and (\ref{77}) are equivalent to the same set of
coupled Eqs. (\ref{4}). Thus, in fact they represent two different forms of the same second-order differential equation for the same energy $E$.
Or, more precisely, the components $\Phi_+$ and $\chi_+$ of the Dirac bispinor correspond to the component $\psi_{+1/2}$ of the Pauli spinor; the
components $\Phi_-$ and $\chi_-$ of the Dirac bispinor correspond to the component $\psi_{-1/2}$ of the Pauli spinor. Thus, unlike the
conventional approach (see, e.g., \cite{Mess}) where the 'small' ('light', in our terms) component $\chi$ is associated with negative energies, in
our approach both the 'heavy' and 'small' quasiparticles have the same energy and move in the same vector potential. Hence both $\Phi$ and $\chi$
describe the {\it particle} states (or both describe the {\it antiparticle} states).

Such states are invariant with respect to the Lorentz transformations and represent a complete set of states of a relativistic particle. The fact
that Eqs. (\ref{4}) possess also solutions with the negative values of $E$ means simply that these equations imply also, in addition to the
particle states, the existence of the antiparticle states. Again, the set of the antiparticle states is invariant under the Lorentz
transformations and, thus, it represents a complete set of states of a relativistic antiparticle.

Of course, the group of transformations of symmetry of the Dirac equation contains also the operation of charge conjugation that transforms the
particle and antiparticle states into each other. But this transformation essentially changes the physical context that determines the properties
of a quantum ensemble (it changes the signs of the external static fields $V$ and $\vec{\cal{A}}$) and hence it transforms one quantum ensemble
into another. The particle and antiparticle states cannot be mixed with each other: a superposition of states with the positive and negative
values of $E$ in (\ref{3}) (and hence Schr\"{o}dinger's version of Zitterbewegung) is prohibited.

All this means that it is sufficient to solve Eqs. (\ref{4}), with the potentials $V$ and $\vec{\cal{A}}$, for a {\it particle} and then apply
these solutions to the corresponding {\it antiparticle} moving under the potentials $-V$ and $-\vec{\cal{A}}$. In doing so, we have to take into
account that for the static electric field, for example, all particle states lie in the region $\epsilon>V_{min}$, where $V_{min}$ is the minimal
value of the scalar potential $V(\vec{r})$ for a given structure.

\section{Scattering a Dirac particle on the potential step}

Our next step is to study the Klein tunneling. Thus, it is sufficient to consider the scattering problem where the vector potential
$\vec{\cal{A}}$ is zero and the scalar potential $V$ depends only on $z$, representing a piecewise constant function: $V(z)=0$ for $z<0$ and
$V(z)=V_0$ for $z>0$; $V_0$ is constant. We will also assume that a particle moves toward the potential step from the left, strictly in
$z$-direction. Since $V_{min}=0$ in this problem, all states lie in the region $\epsilon>0$.

Note that, in this scattering problem, equations for both spin components are separated from each other. Thus, it is sufficient to consider only
the equations for the upper spin. Since Eqs. (\ref{7}) and (\ref{77}) are equivalent, the components $\Phi_+$ and $\chi_+$ are described by the
same second-order differential equation which can be written as
\begin{eqnarray} \label{9}
-\frac{\hbar^2}{2M}\frac{d^2\Psi}{dz^2}+V_0\theta(z)\Psi=\epsilon\Psi,
\end{eqnarray}
where $\theta(z)$ is the Heaviside function. The wave function $\Psi(z;E)$ which represents the pair ($\Phi_+,\chi_+$) (and the pair
($\Phi_-,\chi_-$)) is continuous, at the point where the potential $V(z)$ is discontinuous, together with the function
$\frac{1}{M(\epsilon,V(z))}\frac{d\Psi(z;E)}{dz}$. The corresponding probability density $W$ and the probability current density $J_z$ are
\begin{eqnarray} \label{10}
W=|\Psi|^2+\frac{\hbar^2}{4M^2c^2}\left|\frac{d\Psi}{dz}\right|^2,\ppp J_z=\frac{i\hbar}{2M}\left(\Psi \frac{d\Psi^*}{dz}- \Psi^*
\frac{d\Psi}{dz}\right).
\end{eqnarray}
Note that the components $\Phi_+$ and $\Phi_-$ are described in general by different solutions of Eq. (\ref{9}). As regards the components
$\chi_+$ and $\chi_-$, they are determined by the equality $\chi_{\pm}=\mp\frac{i\hbar}{2Mc}\frac{d\Phi_{\pm}}{dz}$.

Since the effective masses of the heavy and light components are different and constant in the regions $z<0$ and $z>0$, let further
\[M_0=M(\epsilon,0)=m+\mu_0,\ooo \mu_0=\mu(\epsilon,0)=\frac{\epsilon}{2c^2},\ooo M_V=M(\epsilon,V_0)=m+\mu_V,\ooo
\mu_V=\mu(\epsilon,V_0)=\frac{\epsilon-V_0}{2c^2}.\] Then the general solution of Eq. (\ref{9}) in the region $z<0$, where the particle is free
and the effective masses $M$ and $\mu$ of both its components are positive, can be written as follows
\begin{eqnarray} \label{11}
\Psi=A_1 e^{ikz}+B_1 e^{-ikz},\ppp \hbar k=\sqrt{2M_0\epsilon}\equiv 2c\sqrt{\mu_0 M_0};
\end{eqnarray}
$A_1$ and $B_1$ are constants to be determined.

For the region $z>0$ we have
\begin{subequations}
\begin{align}
\Psi=A_2 e^{i\kappa z}+B_2 e^{-i\kappa z},\ppp \hbar\kappa=\beta\sqrt{2M_V(\epsilon-V_0)}\equiv 2\beta c\sqrt{\mu_V M_V}\ppp (\mu_V M_V>0); \label{12a}\\
\Psi=A_2 e^{-\kappa z}+B_2 e^{\kappa z},\ppp \hbar\kappa=\sqrt{2M_V(V_0-\epsilon)}\equiv 2 c\sqrt{-\mu_V M_V}\ppp (\mu_V M_V<0); \label{12b}
\end{align}
\end{subequations}
$\beta=sign(M_V)$, $A_2$ and $B_2$ are arbitrary constants.

Now, when the effective masses of both quasiparticles have opposite signs, when they move under the influence of different potentials, we are
facing a more complex situation which depends on the energy of the quasiparticles. Indeed, for $V_0>2mc^2$ we have the following possibilities:
\begin{itemize}
\item[$\bullet$] When $\epsilon>V_0$ (and hence $\epsilon>V_0-2mc^2$) both heavy and light quasiparticles have the positive
effective masses and both move in the above-barrier regime.
\item[$\bullet$] When $V_0-2mc^2<\epsilon<V_0$ the heavy quasiparticle has the positive effective mass $M_V$ and moves, in the region $z>0$,
in the under-barrier regime -- this spatial region is classically forbidden for it. At the same time $\mu_V<0$ and, thus, the light quasiparticle
behaves in the region $z>0$ like an anti-particle. As a consequence, though $\epsilon>V_0-2mc^2$ as in the above case, the region $z>0$ is now
classically forbidden for the light quasiparticle.
\item[$\bullet$] When $0<\epsilon<V_0-2mc^2$ (the Klein zone) the effective masses of both quasiparticles are negative. As a consequence, they behave
in the region $z>0$ like anti-particles; that is, this region is classically accessible for them.
\end{itemize}

As is seen from this analysis, despite the different effective masses, the heavy and light quasiparticles behave equally in all the energy
intervals. Note that when $\mu_V M_V<0$ the region $z>0$ is classically forbidden for both quasiparticles; otherwise it is classically accessible
for them.

\subsection{Total reflection}

Let us first consider the case when $\mu_V M_V<0$. This takes plays when $V_0-2mc^2<\epsilon<V_0$ for $V_0>2mc^2$; otherwise, this condition can
be written as $0<\epsilon<V_0$. In both cases, $M_V>0$ but $\mu_V<0$.

Of course, since $\Psi$ should be everywhere bounded, $B_2=0$. Then, matching the solutions (\ref{11}) and (\ref{12b}) at the point $z=0$, with
making use of the continuity conditions
\begin{eqnarray} \label{13}
\Psi|_{z=-0}=\Psi|_{z=+0},\ppp \frac{1}{M_0}\frac{d\Psi}{dz}\Big|_{z=-0}=\frac{1}{M_V}\frac{d\Psi}{dz}\Big|_{z=+0}
\end{eqnarray}
we find the constants $B_1$ and $A_2$:
\begin{eqnarray} \label{14}
B_1=\frac{\tilde{k}-i\tilde{\kappa}}{\tilde{k}+i\tilde{\kappa}} A_1,\ooo A_2= \frac{2\tilde{k}}{\tilde{k}+i\tilde{\kappa}}A_1;\ppp
\tilde{k}=\frac{k}{M_0}\equiv \frac{2c}{\hbar}\sqrt{\frac{\mu_0}{M_0}},\ooo \tilde{\kappa}=\frac{\kappa}{M_V}\equiv
\frac{2c}{\hbar}\sqrt{-\frac{\mu_V}{M_V}}.
\end{eqnarray}

As it was expected, $|B_1|=|A_1|$: Exps. (\ref{11}) and (\ref{12b}), with the constants $B_1$ and $A_2$ (\ref{14}), represent a standing wave. In
this case the probability current density $J_z$ (see expression (\ref{10})) is zero; the incident and reflected flows coincide with each other --
total reflection.

\subsection{The Klein tunneling and passage of a particle above the potential step}

Let now $\mu_V M_V>0$. This takes place in the following two cases: when $\epsilon>V_0$ -- the passage of a particle above the potential step;
when $0<\epsilon<V_0-2mc^2$ -- the Klein tunneling (this implies that $V_0>2mc^2$).

Matching the solutions (\ref{11}) and (\ref{12a}) at the point $z=0$, with making use of the continuity conditions (\ref{13}), we obtain
\begin{eqnarray} \label{16}
\left(\begin{array}{cc} A_1 \\ B_1 \end{array}\right)=\alpha\vec{Y}\left(\begin{array}{cc} A_2 \\ B_2 \end{array}\right);\ppp
\vec{Y}=\left(\begin{array}{cc} q & p \\ p & q
\end{array}\right),\ooo q=\frac{1}{\sqrt{T}}=\theta_+,\ooo p=\sqrt{\frac{R}{T}}=\theta_-; \\
\theta_\pm = \frac{1}{2}\left(\alpha^{-1}\pm\alpha\right),\ppp \alpha=\sqrt{\frac{\tilde{\kappa}}{\tilde{k}}},\ppp
\tilde{\kappa}=\frac{\kappa}{M_V}\equiv \frac{2c}{\hbar}\sqrt{\frac{\mu_V}{M_V}};
\end{eqnarray}
here $\vec{Y}$ is the transfer matrix of the potential step; $T$ and $R$ are the transmission and reflection coefficients, respectively; $R=1-T$;
note that $\alpha<1$. Note, $T=1$ for a particle with $m=0$.

Since a particle source is located on the left of the step the wave $B_2 e^{-i\kappa z}$, associated with the negative current density, must be
discarded: $B_2=0$. (In the Klein zone this wave has a positive phase velocity ($-\kappa>0$) and sometimes namely this wave is erroneously
considered as an essential.) As a consequence,
\begin{eqnarray*}
\Psi=A_1[\exp(ikz)+\sqrt{R}\exp(-ikz)]\ooo (z<0); \ppp \Psi=A_1\alpha^{-1}\sqrt{T}\exp(i\kappa z)\ooo  (z>0).
\end{eqnarray*}
Note that the probability density $W_{tr}$ in the region $z>0$ as well as the (total) probability current density $J_z$ (see Exps. (\ref{10})) are
\begin{eqnarray*}
W_{tr}=\left(1+\frac{\hbar^2\tilde{\kappa}^2}{4c^2}\right)|A_2|^2\equiv \left(1+\frac{\mu_V}{M_V}\right)|A_2|^2,\ppp
J_z=\hbar\tilde{\kappa}|A_2|^2\equiv 2c\sqrt{\frac{\mu_V}{M_V}}\ooa |A_2|^2.
\end{eqnarray*}
Thus, the 'flow' velocity $v_{flow}=J_z/W_{tr}$ in the region $z>0$ is
\begin{eqnarray*}
v_{flow}=2c\frac{\sqrt{\mu_V M_V}}{|M_V+\mu_V|}\equiv \frac{\hbar \kappa}{\cal{M}}.
\end{eqnarray*}
As is seen, $0<v_{flow}<c$ both for $\epsilon>V_0$ and in the Klein zone. The only peculiarity of the Klein zone is that now the flow and phase
velocities of the wave $A_2\exp(i\kappa z)$ have the opposite signs. This is so because the repulsive potential $V$ becomes attractive, in the
Klein zone, for both the 'heavy' and the 'light' components; their effective masses $M_V$ and $\mu_V$ are negative in this zone.

Here it is also important to note that both terms in Exp. (\ref{10}) for $W$ -- the first one that corresponds to the 'heavy' component of the
Dirac bispinor, as well as the second one that corresponds to its 'light' component -- are necessary in order to guarantee the fulfillment of the
inequality $v_{flow}<c$.

\section{Discussion and conclusion} \label{concl}

Through the example of a Dirac particle with a given energy $E$, moving orthogonally to the layers of a spatial structure described by the static
scalar potential $V$ and the vector potential $\vec{\cal{A}}$, it is shown that the set of two coupled first-order differential equations for the
'large' ($\Phi$) and 'small' ($\chi$) components of the Dirac bispinor can be presented in the following two equivalent forms: (\i) in the form of
the Pauli equation for the component $\Phi$ that describes the quantum dynamics of a 'heavy' quasiparticle with the effective mass $M$, in these
fields; (\i\i) in the form of the Pauli equation for the component $\chi$ that describes the quantum dynamics, in the same vector potential but in
the scalar potential $V-2mc^2$, of a 'light' quasiparticle with the effective mass $\mu$.

This means that by our approach the ensemble of Dirac particles with the energy $E$, moving in the four-dimensional space-time under the influence
of the scalar potential $V$ and vector potential $\vec{\cal{A}}$, consists of two subensembles of 'heavy' and 'light' Pauli quasiparticles with
the same energy $E$, moving in the Euclidian three-dimensional space under the same vector potential $\vec{\cal{A}}$. As regards the scalar
potential $V$, the 'heavy' quasiparticle 'sees' it as it stands, while the 'light' quasiparticle 'sees' the reduced potential $V-2mc^2$, rather
than $V$. These quasiparticles can be converted into each other in the course of scattering (see the second term in Eq. (\ref{7})): only the total
number of Dirac particles in the ensemble is conserved. The effective mass of each Pauli quasiparticle contains information about the Lorentzian
symmetry of the four-dimensional space-time: $M$ and $\mu$ are dynamical rather than inertial or gravitational masses of the Dirac particle;
$\mu/M\to 1$ when $\varepsilon/mc^2\to\infty$.

In fact, this approach says once more that the four-dimensional space-time is not empty. The space is filled with a physical vacuum, and of
importance is to reveal the role of this vacuum in 'forming' the effective masses of the 'heavy' and 'light' internal degrees of freedom of the
Dirac particle. This is not a prerogative of quantum mechanics which describes the nature on the statistical level. Rather it is the task of QED
(quantum field theory) which should be treated as a sub-quantum theory.

Since the Dirac bispinor like the Pauli spinor (two-component Schr\"{o}dinge wave function) describes a quantum particle on the statistical level,
there is no reason to believe (at this level) that the Dirac particle is a physical object consisting of these two quasiparticles. It is rather
the {\it ensemble} of Dirac particles with the 'mass' $(E-V)/c^2$ that consists of two subensembles of quasiparticles with the effective masses
$M$ and $\mu$, such that $M+\mu=(E-V)/c^2$. The Dirac particle with the energy $E$ can move {\it either} as a heavy quasiparticle {\it or} as a
light quasiparticle, with the probabilities $|\Phi|^2$ and $|\chi|^2$, respectively. Any averaging is allowed only for both subensembles of
quasiparticles.

Contrary to the conventional approach \cite{Mess}, in our approach the 'small' component $\chi$ remains essential even in the non-relativistic
limit: firstly, both the 'large' ('heavy') and 'small' ('light') components are equally important in the expression (\ref{2}) for the probability
current density; secondly, both the components are equally important for transforming the system (\ref{4}) of the first-order differential
equations into the equivalent second-order differential (generalized) Pauli equations (\ref{7}) and (\ref{77}). Besides, due to the interplay
between the heavy and light components, the coordinate uncertainty of a Dirac particle with the mass $m$, unlike a Schr\"{o}dinger particle with
the same mass, has a lower limit (see below).

Thus, according to our approach there is a close relationship between the Lorentzian Dirac's dynamics and Euclidean Schr\"{o}dinger's dynamics.
Namely, {\it solving} the stationary Dirac equation is reduced to {\it solving} the generalized Pauli equation for a quasiparticle with the
spatially-dependent effective mass. Moreover, in the non-relativistic limit this generalized equation coincides with the Pauli equation. On the
one hand, this means that Dirac theory is indeed a quantum theory of single fermions. On the other hand, this means that (Schr\"{o}dinger's)
quantum mechanics is compatible with special relativity.

However, this does not at all mean that the Dirac {\it dynamics} of a particle with $m\neq 0$ is reduced to the generalized Schr\"{o}dinger {\it
dynamics}. They do not coincide even in the non-relativistic limit, because the former unlike the latter implies the existence of a lower limit of
the coordinate uncertainty for this particle. Indeed, let us consider a free particle in one dimension and let $L_M$ be the size of the space
region, where the probability density $|\Psi|^2$ in the expression from (\ref{10}) for $W$ is significantly different from zero, and $\bar{P}$ be
the average value of $|\Psi|^2$ in this region. Then, taking into account that $M^{-1}<m^{-1}$ for a free Dirac particle, we can estimate its
coordinate uncertainty $\nabla z$ as follows,
$$\int_{-\infty}^\infty W(z)dz\stackrel{<}{\sim}
\bar{P}\left(1+\frac{\hbar^2}{4m^2c^2L_M^2}\right)L_M=\bar{P}\left(L_M+\frac{ \lcom_C^2}{4L_M}\right)\equiv \bar{P}\nabla z;\ooo
\lcom_C=\frac{\hbar}{mc}=\frac{\lambda_C}{2\pi};$$ $\lambda_C$ is the Compton wavelength of a Dirac particle. From here it follows that $\nabla
z\stackrel{>}{\sim} \lcom_C$ and $\bar{P}\stackrel{<}{\sim} \lcom_C^{-1}$, where their extremum is reached at $L_M=\lcom_C/2$. Let $L_\mu$ be the
size of the localization region of the light component: $L_\mu=\lcom_C^2/4L_M$. Then we have $L_M+L_\mu\stackrel{>}{\sim} \lcom_C$ and $L_M\cdot
L_\mu=\lcom_C^2/4$.

This analysis not only replays the known estimations of the confinement limit for a free Dirac particle with $m\neq 0$ (see, e.g.,
\cite{Dod,Una,Che}) but it also shows that this limit arises due to the existence of the light component in the quantum ensemble of this particle.
Note also that this limit exactly coincides with the zitter radius obtained in \cite{Hest} for a free particle. That is, our approach indirectly
substantiates the (pre-quantum) Hestenes Zitterbewegung model (however, we think that, on the statistical level, a pre-quantum model of a Dirac
particle must exactly reproduce the Dirac dynamics).

To make the same estimation for a Dirac particle with $m=0$, let us consider this particle in the localized state in which its average energy is
$\bar{\epsilon}$. Then, using the notations $\bar{P}$, $L_M$ and $L_\mu$, we obtain
$$\int_{-\infty}^\infty W(z)dz \approx \bar{P}\left[1+\left(\frac{\hbar}{2M(\bar{\epsilon})cL_M}\right)^2\right]L_M=\bar{P}\left(L+\frac{
\hbar^2c^2}{\bar{\epsilon}^2L_M}\right)\equiv \bar{P}\nabla z;$$ Thus, for a Dirac particle with $m=0$ we obtain $\nabla z\stackrel{>}{\sim}
2\hbar c/\bar{\epsilon}$ and $\bar{P}\stackrel{<}{\sim} \bar{\epsilon}/2\hbar c$; their extremum is reached at $L_M=L_\mu=\hbar c/\bar{\epsilon}$.
In the general case $L_M+L_\mu\stackrel{>}{\sim} 2\hbar c/\bar{\epsilon}$ and $L_M\cdot L_\mu= \hbar^2c^2/\bar{\epsilon}^2$.

Apart from the Hestenes Zitterbewegung, there is also enough room in our approach for the Schr\"{o}dinger version of this phenomenon (but now this
version does not imply a mixing of the particle and antiparticle states). An interesting task is to study the Dirac dynamics of wave packets
consisting of the stationary solutions that correspond to energies from the Klein zone and the above-barrier interval $\epsilon>V_0$. The intrigue
lies in the fact that particles with energies $\epsilon_1=V_0-2mc^2-\Delta\epsilon>0$ and $\epsilon_2=V_0+\Delta\epsilon$ possess the same flow
velocity, while $\epsilon_2-\epsilon_1>2mc^2$. This means that in this case the actual zitter frequency can be much less than the expected value
$2mc^2/\hbar$.

In order to search, within our approach, for realistic physical conditions which could be suitable for experimental observations of
Zitterbewegung, it is necessary to study, at least in one dimension, the temporal dynamics of the probability density $<\Phi|\Phi>+<\chi|\chi>$
and the average position $<\Phi|\hat{z}|\Phi>+<\chi|\hat{z}|\chi>$; $\hat{z}$ is the position operator. In particular, it is important to study
the role of the 'heavy' and 'light' components in their dynamics. We hope that such studies help to shed light on the existing discrepancy between
researchers (see, e.g., \cite{Ger,Con}) with respect to the reality of Zitterbewegung.

Regarding the Klein paradox, its old version where the incident flow of particles is less than the outgoing flows should be considered as a result
of an incorrect statement of the scattering problem for the Klein zone (it is incorrect to set in (\ref{12b}) $A_2=0$ and $B_2\neq 0$). In the
correct statement of this problem the flow of incident particles is always equal to the sum of the absolute values of the outgoing flows. In this
case the transmission coefficient for a Dirac particle scattering on the strong potential step, with the energy in the Klein zone, is not zero. By
our approach this takes place because both quasiparticles have, in this zone, negative effective masses: a repulsive potential acts on them as an
attractive potential.

Of course, this article is only the first step to substantiate the Dirac equation as a quantum mechanical equation for a single relativistic
fermion. Among the immediate tasks we see the study of the {\it temporal aspects} of the Dirac quantum dynamics. In our opinion, this can help to
observe indirectly the individual dynamics of the heavy and light components $\Phi$ and $\chi$ (and to indirectly measure their effective masses).
Besides, an important task is to simulate, on the basis of this approach, electronic properties of graphene. There have been obtained a lot of
experimental and/or theoretical results on the Klein tunneling (see \cite{Been}) and Zitterbewegung effect (see, e.g., \cite{Zaw,Dav}) in this
material, and it is important to rethink these results on the basis of a unified approach.

In this regard, it is useful to dwell on one feature of this approach which is important for studying the electron transport in heterostructures
with layers of graphene. It concerns the fact that the rest mass $m$ of a Dirac particle is constant. At the same time the (non-relativistic)
effective mass $m^*$ of a Bloch electron varies in a heterostructure during the transition from one layer to another. Thus, in order to describe
this peculiarity of a Bloch electron, the presented approach must be generalized onto the Dirac equation that contains, in addition to the static
electric scalar potential, the (piecewise-continuous) Lorentz-scalar potential.

And yet, the boundary conditions for the heavy and light components $\Phi$ and $\chi$ in our approach are unique because they are determined by
the very form of the Dirac equation. At the same time there are three different types of sharp boundaries in heterostructures with graphene. They
lead to the different electron properties in these structures \cite{Zar} and hence it is important to elucidate the question of the conformity of
these types of boundaries with the boundary conditions dictated by the Dirac equation.

It is also important to stress that this approach opens the possibility to apply the mathematical methods of solving the stationary
Schr\"{o}dinger equation to the Dirac equation. In particular, this concerns the well-known transfer-matrix approach which is suitable for solving
the Schr\"{o}dinger equation with piecewise constant effective mass and potential function.

\section{Acknowledgments}

First of all I would like to thank Prof. I. L. Buchbinder for his useful critical remarks on the first version of the paper. I also thank Prof. V.
G. Bagrov, Prof. V. A. Bordovitsyn and Prof. G. F. Karavaev for useful discussions on this subject. Finally, I want to express my deep gratitude
to Reviewers for their helpful remarks and questions. This work was supported in part by the Programm of supporting the leading scientific schools
of RF (grant No 88.2014.2) for partial support of this work.


\end{document}